\documentclass[journal=nanoletters,manuscript=article,layout=twocolumn]{achemso}

\setkeys{acs}{keywords = true}

\usepackage[utf8]{inputenc}
\usepackage[OT4]{fontenc}
\usepackage{amsmath}
\usepackage{xcolor}
\usepackage{pdfpages}

\newcommand*{\citen}[1]{%
	\begingroup
	\romannumeral-`\x 
	\setcitestyle{numbers}%
	\cite{#1}%
	\endgroup   
}

\author{Ievgeniia Chaban}
\affiliation{Department of Chemistry, Massachusetts Institute of Technology, Cambridge, MA 02139, USA}
\author{Radoslaw Deska}
\affiliation{Advanced Materials Engineering and Modelling Group, Wroclaw University of Science and Technology, PL-50370 Wroclaw, Poland}
\author{Gael Privault}
\affiliation{Institut de Physique de Rennes, UMR CNRS 6251, Universit\'e Rennes 1, 35042 Rennes Cedex, France}
\author{Elzbieta Trzop}
\affiliation{Institut de Physique de Rennes, UMR CNRS 6251, Universit\'e Rennes 1, 35042 Rennes Cedex, France}
\author{Maciej Lorenc}
\email{maciej.lorenc@univ-rennes1.fr}
\affiliation{Institut de Physique de Rennes, UMR CNRS 6251, Universit\'e Rennes 1, 35042 Rennes Cedex, France}
\author{Steven E. Kooi}
\affiliation{Institute for Soldier Nanotechnologies, Massachusetts Institute of Technology, Cambridge, MA 02139, USA}
\author{Keith A. Nelson}
\affiliation{Department of Chemistry, Massachusetts Institute of Technology, Cambridge, MA 02139, USA}
\author{Marek Samoc}
\affiliation{Advanced Materials Engineering and Modelling Group, Wroclaw University of Science and Technology, PL-50370 Wroclaw, Poland}
\author{Katarzyna Matczyszyn}
\email{katarzyna.matczyszyn@pwr.edu.pl}
\affiliation{Advanced Materials Engineering and Modelling Group, Wroclaw University of Science and Technology, PL-50370 Wroclaw, Poland}
\author{Thomas Pezeril}
\affiliation{Department of Chemistry, Massachusetts Institute of Technology, Cambridge, MA 02139, USA}
\alsoaffiliation{Institut de Physique de Rennes, UMR CNRS 6251, Universit\'e Rennes 1, 35042 Rennes Cedex, France}
\email{pezeril@mit.edu}

\title{Nonlinear optical absorption in nanoscale films revealed through ultrafast acoustics}


\begin{document}


\begin{abstract}


{\color{black}Herein we describe a novel spinning pump-probe photoacoustic technique developed to study nonlinear absorption in thin films. As a test case, an organic polycrystalline thin film of quinacridone, a well-known pigment, with a thickness in the tens of nanometers range, is excited by a femtosecond laser pulse which generates a time-domain Brillouin scattering signal. This signal is directly related to the strain wave launched from the film into the substrate and can be used to quantitatively extract the nonlinear optical absorption properties of the film itself. Quinacridone exhibits both quadratic and cubic laser fluence dependence regimes which we show to correspond to two- and three-photon absorption processes. This technique can be broadly applied to materials that are difficult or impossible to characterize with conventional transmittance-based measurements including materials at the nanoscale, prone to laser damage, with very weak nonlinear properties, opaque or highly scattering.}

\end{abstract}

\maketitle


Nonlinear optical (NLO) absorption coefficients, key sample parameters for applications such as optical limiting, saturable optical absorbers, and photothermal cancer therapy, are traditionally obtained from Z-scan optical transmission based measurements\cite{Sheik} or, where possible, through luminescence based techniques. Due to its simplicity, the Z-scan technique is widely used, however, for many nanomaterials or nanostructures, this technique is not suitable as it often requires relatively thick samples (typically in the micron to millimeter range), in order to reach a decent sensitivity for the detection of transmittance changes. Nanoparticle samples often need to be studied as liquid or solid suspensions to reach the suitable thickness and highly scattering or opaque media cannot be investigated. This largely restricts the applicability of the Z-scan technique and has lead us to search for alternative ways to detect nonlinear absorption to bridge the experimental gap. In the present paper, we highlight the benefits of a new GHz photoacoustic time-domain Brillouin scattering based technique, which is very well suited for the study of nanoscale samples of many different sorts and allows for the extraction of nonlinear optical absorption parameters.

Photoacoustic spectroscopy\cite{Fayer,West,Haisch} and photoacoustic imaging\cite{Hu} are well-established techniques in which the conversion of electromagnetic energy into mechanical energy is exploited. This energy conversion process involves the absorption of a pulse of light in a material resulting in localized heating and simultaneous mechanical strain loading upon thermal dilatation. In fact, photoacoustic detection in the MHz ultrasonic range has already been implemented in the standard Z-scan technique and has proven to be useful for the determination of the linear as well as the nonlinear optical absorption coefficients while being not susceptible to other loss processes including light scattering\cite{Chandra,Chantharasupawong,Kislyakov}. Unfortunately, these modified Z-scan techniques are still not suitable for measurements of nanoscale samples for which GHz ultrasonic frequencies necessitate using femtosecond pulses for the optical excitation and detection of the ultrasound signals.

Recently, Zeuschner et al.\cite{Zeuschner} demonstrated the use of ultrafast acoustic strains to determine the NLO absorption properties of crystalline thin films. In their experiments, femtosecond laser pulses were used to excite ultrafast strains and a femtosecond X-ray probe was used to capture time-resolved X-ray diffraction peak modulation due to change of lattice constants produced by the induced strains. Even though a careful analysis of the laser-induced thermal strain in the crystalline thin-films gave insight into the linear as well as the nonlinear optical absorption processes, that technique requires very specialized instrumentation and is limited to crystalline samples only. On the contrary, as it will be shown in the present paper, our technique is based on instrumentation available in many laser laboratories and is applicable to a wide range of ultra-thin samples. We employ a method adapted from picosecond interferometry or time-domain Brillouin scattering (TDBS)\cite{Thomsen, Maris1991,Matsuda2015}, based on a conventional femtosecond near-infrared pump - visible probe scheme, and demonstrate that the probing of GHz frequency acoustic waves in a supporting substrate, on which the material of interest is grown or deposited, can be used to characterize nonlinear absorption processes. We illustrate the capability and unprecedented sensitivity of the new technique for the quantitative evaluation of the relevant NLO parameters of a soft, fragile and highly absorbing organic compound.

\section{Methods}

\begin{figure}[t!]
\centering
\includegraphics[width=8cm]{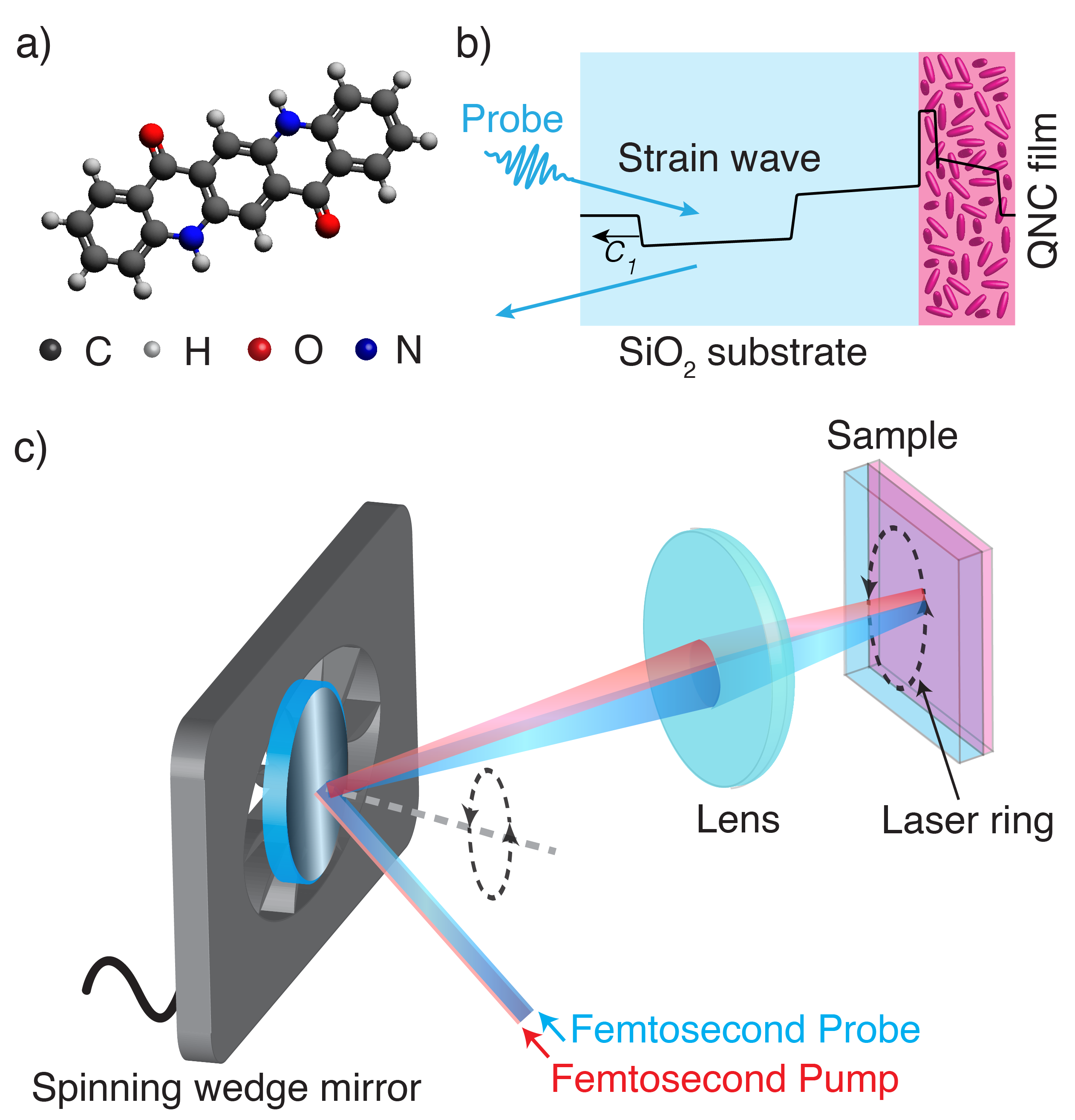}
\caption[Fig1]{(Color online) (a) Illustration of the experimental pump-probe configuration including a spinning wedge mirror, that is used to scan the laser beams over the sample surface at high speed in order to avoid laser damage. (b) Schematic view of the QNC molecule. (c) Sketch of the experiment. The QNC film absorbs the femtosecond pump light resulting in thermoelastic expansion which launches a mechanical strain wave that gets transmitted into the glass substrate. A subsequent femtosecond probe is scattered by the strain wave propagating in the substrate at the speed of sound $c_1$, and is used to measure the time-domain acoustically-induced changes in refractive index of the film.}
\label{fig1}
\end{figure}

Thin-films samples were prepared by physical vapor deposition (PVD) of an organic compound, 5,12-dihydroquinolino[2,3-b]acridine-7,14-dione (C$_{20}$H$_{12}$N$_{2}$O$_{2}$), commonly known as quinacridone (QNC), a widely used magenta pigment. The structure of an individual QNC molecule is sketched in Fig.~\ref{fig1}(a). The results detailed in this paper were obtained from a QNC film of 38~nm thickness, deposited on a flat 1~mm thick glass substrate, see Fig.~\ref{fig1}(b). Note that the sample requirements are quite moderate. As in any time-domain Brillouin scattering measurements, the sample layer should be of optical quality, which is not a very high constraint and of easy reach for many standard deposition techniques (sputtering, spin-coating, vapor deposition…). As in \cite{Pezeril2009, Klieber2012, Chaban2020, Klieber}, the sample could even be an ultra-thin liquid film. The only constraint on the sample is that it should be in contact with a transparent substrate (e.g. preferentially a glass plate for which the photoacoustic coefficients are already well-known at the probe wavelength\cite{matsuda}) and its thickness should not be significantly bigger than, typically, one micron, otherwise the laser excited strain wave may become excessively damped before being transmitted into the substrate.

In order to surpass the limitations of the Z-scan technique, see Supplemental for details on the inconclusive Z-scan measurements performed on the exact same sample, we have implemented a novel spinning ultrafast photoacoustic technique.  This laser-spinning technique sketched in Fig.~\ref{fig1}(c), that would even mimic single-shot pump-probe experiments with a faster spinning mirror such as in \cite{Lu}, is extremely useful to avoid cumulative laser damage of the sample \cite{Vaudel} and its use was pivotal for obtaining results described in this paper.

When the QNC film absorbs energy from the pump laser pulse, irrespective whether the energy absorption occurs due single- or multi-photon absorption, an ultrafast rise in local temperature results in thermal dilatation and launches a propagating acoustic wavepacket throughout the thickness of the QNC film. This acoustic wavepacket travels back and forth in the QNC film and simultaneously is partially transmitted into the glass substrate. The acoustic wavepacket is then optically detected by the time-delayed probe pulse through TDBS \cite{Maris1991}. The portion of the probe beam reflected at the sample interface is scattered by the acoustic wave and is then directed and focused onto a photodiode coupled to a lock-in amplifier that is synchronized to the pump modulation frequency in order to measure transient differential reflectivity $\Delta$R(t)/R as a function of time delay between pump and probe pulses. 

\section{Results}

\begin{figure}[t!]
\centering
\includegraphics[width=8cm]{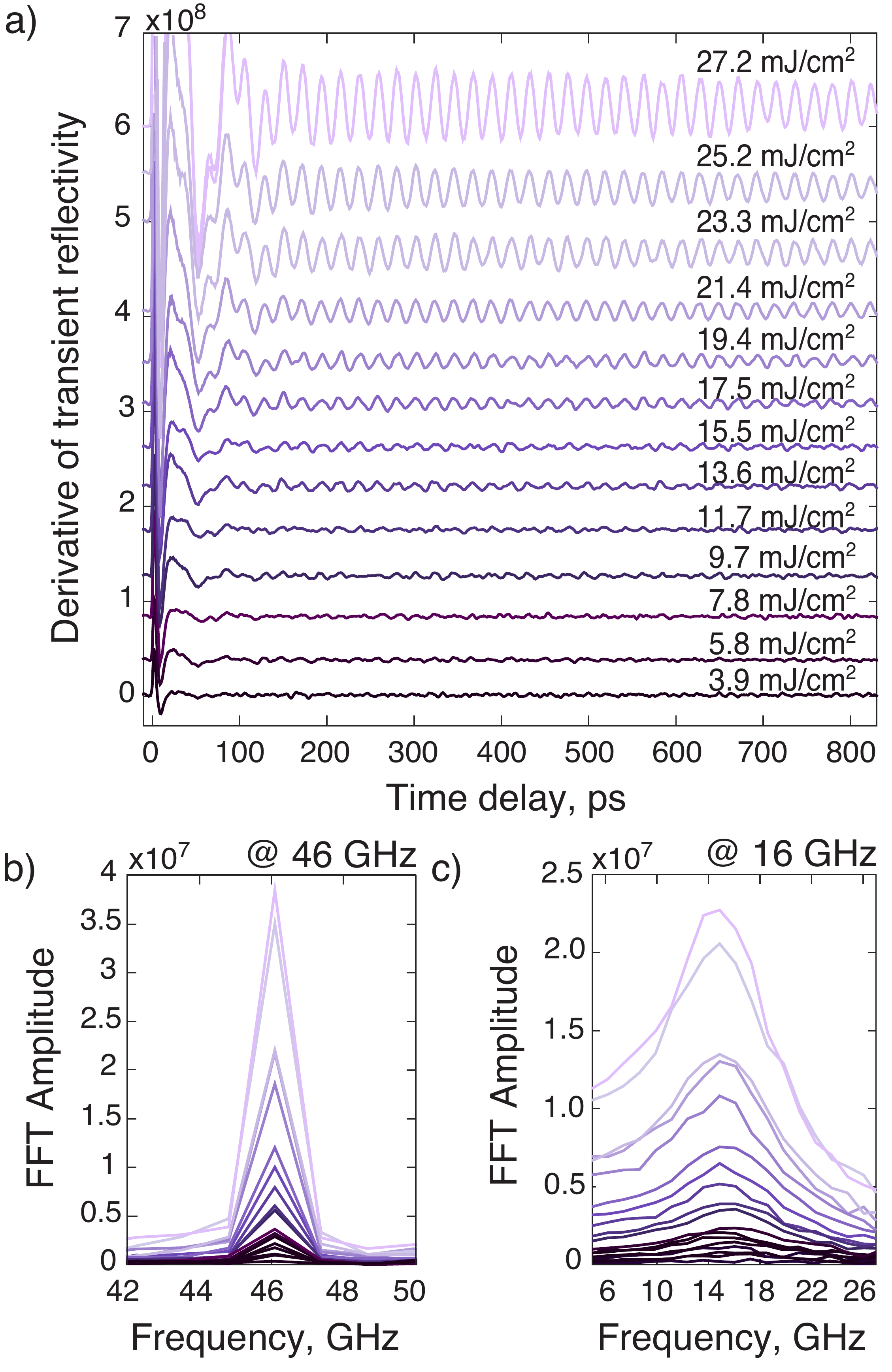}
\caption[Fig2]{(Color online) (a) Time derivative of the recorded transient reflectivity signals obtained at different pump fluences. (b) Fourier amplitudes of the 46~GHz Brillouin
oscillations in glass. (c) Fourier amplitudes of the 16~GHz acoustic resonance. }
\label{fig2}
\end{figure}  

In order to track the nonlinear excitation of the propagating strains due to nonlinear optical absorption in the QNC film, we performed a series of TDBS measurements over a wide range of pump fluences, while keeping the probe power constant at 160~$\mu$W, corresponding to a fluence of about {\color{black}1.2~mJ/cm$^2$}, throughout the measurements. The recorded TDBS data taken at many different pump fluences are shown in Fig.~\ref{fig2} (a). As shown in Fig.~\ref{fig2}(a) the derivative of transient reflectivity for time delays $>$100~ps, propagation of the acoustic wavepacket in the transparent glass substrate leads to long lasting oscillations in the TDBS signal. The frequency $\nu_1$ of these oscillations, often termed Brillouin frequency oscillations, is related to the velocity of the acoustic waves $c_1$ and to the index of refraction $n$ of the glass substrate, in the well-known form, 
\begin{equation}
\label{Brillouin}
\nu_1 = \frac{2 n c_1}{\lambda},
\end{equation}
where $\lambda$ is the probe wavelength. In addition to the high frequency Brillouin oscillations detected in the glass substrate, we note in Fig.~\ref{fig2}(a) from 0 to 150~ps, a lower frequency that corresponds to light scattering from the portion of the acoustic strain confined in the QNC film itself. We model the QNC acousto-optical contribution to the signal as a sinusoidal oscillation at a frequency $\nu_2$ which corresponds to the acoustic resonance of the film of thickness $H$ and acoustic velocity $c_2$, in the form $\nu_2 = c_2/4H$. In total, the normalized TDBS signal can be described by the functional form
\begin{eqnarray}
\label{Reflectivity}
\frac{\Delta {\text{R}}}{\text{R}}\,\ \equiv &&A_1 \exp (-\Gamma_1 t) \cos {(2\pi \nu_1 t + \phi_1)} \,\,\,\,\,\,\, \\
&+& A_2 \exp (-\Gamma_2 t) \cos {(2\pi \nu_2 t + \phi_2)} \nonumber
\end{eqnarray}
where $A_i$ is the amplitude at zero time delay, $\Gamma_i$ is the attenuation coefficient, $\nu_1$ is the Brillouin frequency of the glass substrate, $\nu_2$ is the resonance frequency of the film and $\phi_i$ is the phase that takes into account the excitation process and the acoustic travel time across the film. As is seen in the FFT data analysis of Fig.~\ref{fig2}(b) and (c), the Brillouin frequency $\nu_1$ is centered around 46~GHz, which matches perfectly the expected Brillouin frequency in glass at the probe wavelength \cite{Pezeril2009,Klieber2012, Klieber}, while the lower frequency oscillation $\nu_2$ is centered around 16~GHz. Note that the FWHM of these two distinct frequency peaks are notably broader for the lower frequency component than for the higher frequency one. This is expected because the energy of the acoustic resonance vanishes quickly due to acoustic transmission into the substrate. In contrast, {\color{black}damping of the signal at the Brillouin frequency} is only due to acoustic attenuation in the glass substrate where acoustic waves can propagate over long micrometric distances, corresponding to hundreds of picoseconds, before being significantly damped. As seen in Fig.~\ref{fig2}(b) and (c) and as expected, the amplitudes of both oscillations become more prominent as the pump fluence increases. This is a general trend related to the laser-excitation mechanism that leads to increase of the strain amplitude when the pump fluence increases. In the case of the most general thermoelastic excitation mechanism, that is common in absorptive materials, the signal amplitude is proportional to the strain amplitude and the latter increases linearly with the pump fluence. The Gruneisen parameter is typically used to connect the absorbed laser energy to the lattice strain, and it is valid independently of how photons are absorbed. One can expect the strain amplitude to be proportional to the absorption rate whether the absorption has linear or nonlinear character \cite{Zeuschner}. Since the QNC film is not a single photon absorber at the excitation wavelength of 772 nm, see absorption spectrum in the Supplemental, the NLO absorption is expected to dominate. The below discussion of the results is based on this principle.   

\section{Discussion}

\begin{figure}[!t]
\centering
\includegraphics[width=8cm]{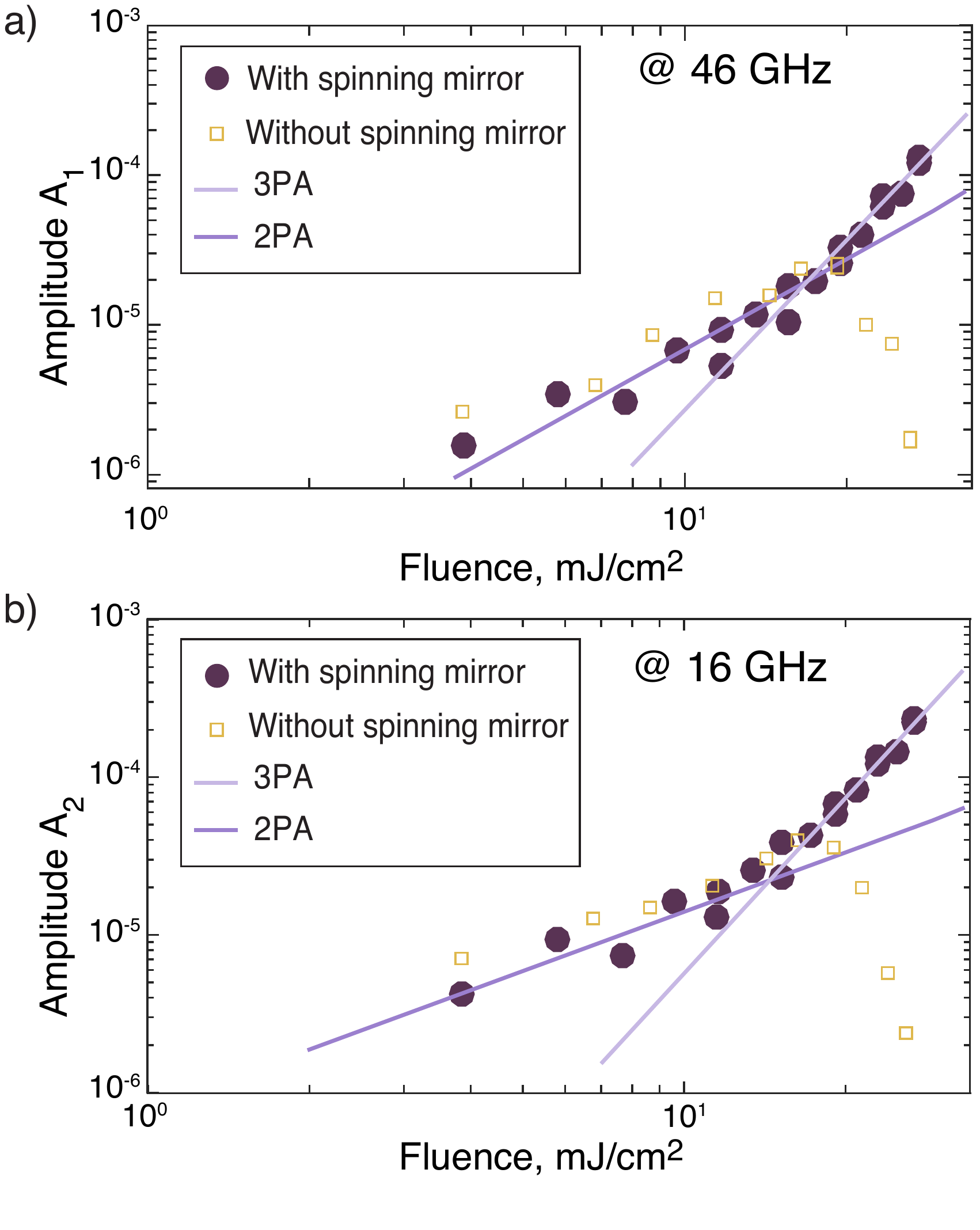}
\caption[Fig3]{(Color online) Fluence dependence of (a) the 46~GHz Brillouin oscillations in glass and (b) 16~GHz oscillations. Note that the fluence $F_0$ displayed on the graphs is the fluence measured in air. }
\label{fig3}
\end{figure}  

The experimental results depicted in Fig.~\ref{fig2} are consistent with the expected amplitude increase with increase in fluence. Further analyses indicate a highly nonlinear dependence. Fig.~\ref{fig3}(a) and (b) show fluence dependent amplitude $A_1$ of the 46~GHz Brillouin oscillations and $A_2$ of the 16~GHz resonance oscillations. Different fluence dependence regimes were probed, including the quadratic dependence that is expected for two-photon absorption and the cubic dependence that is expected for three-photon absorption. While the data up to about 15~mJ/cm$^2$ (which corresponds to an intensity of about 100~GW/cm$^2$) can be reasonably fit with a quadratic dependence, this does not hold at higher fluences where a cubic dependence of the TDBS signal amplitude fits our experimental data much better. It is noted that the extension of the range of usable fluences above 15~mJ/cm$^2$ was only possible with the use of the spinning mirror, otherwise laser damage of the sample would occur, see Fig.~\ref{fig3}. Sample damage encountered during the use of the spinning technique is expected to come from mechanisms such as photoionization and bond breaking while in the non-spinning technique, damage is likely a result of cumulative sample heating by high repetition rate pulses. To ensure the reproducibility of the data shown in Fig.~\ref{fig3}, the experiments were performed by cycling the fluence back and forth, up to a maximum fluence about 20\% below the laser damage threshold.

\begin{figure}[!t]
\centering
\includegraphics[width=8cm]{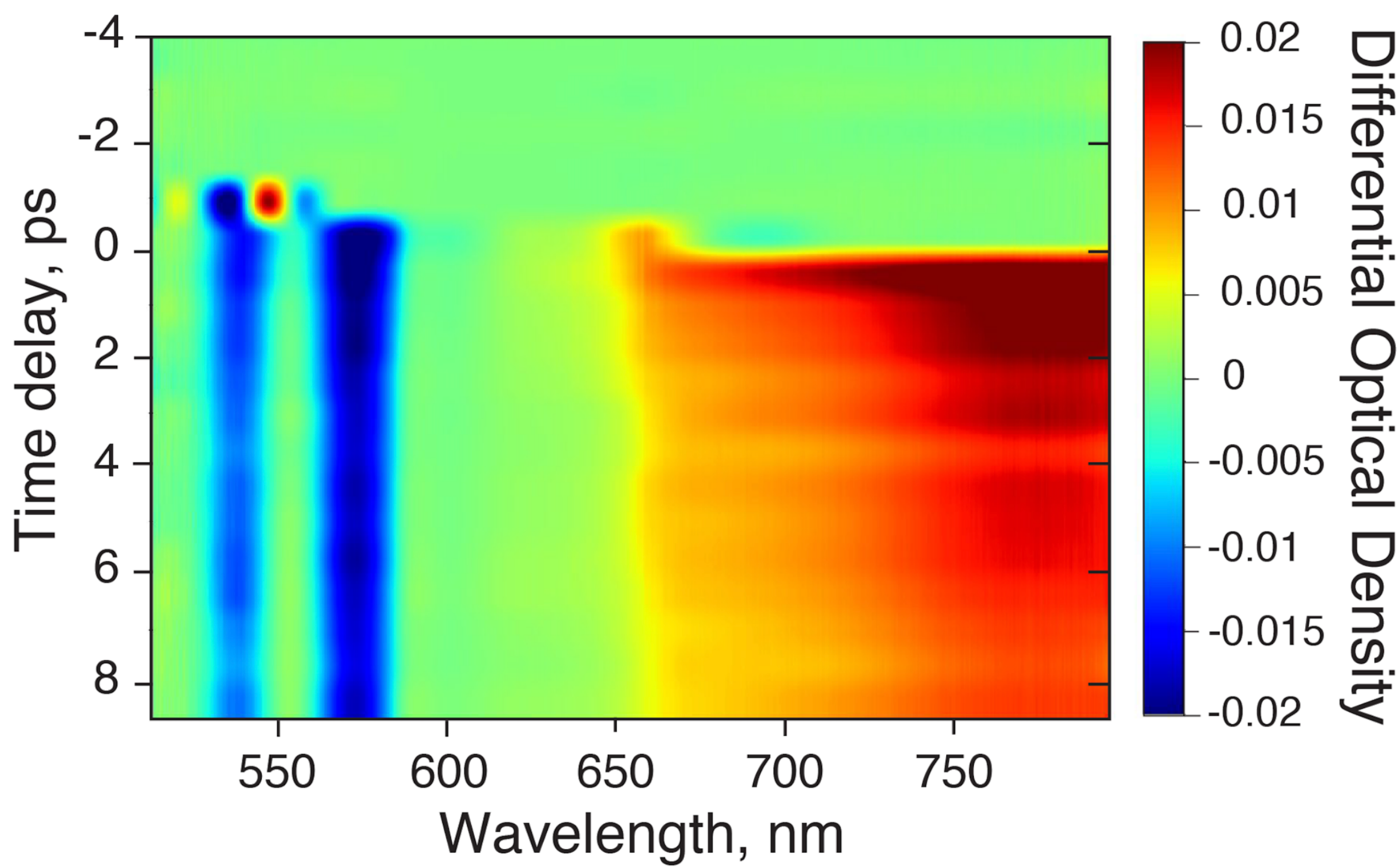}
\caption[Fig4]{(Color online) Transient linear absorption of the QNC film at ultrafast timescale for a 420~nm pump: red - excited state absorption, blue - ground state depopulation. Spectra around time zero were not corrected for chirp. }
\label{fig4}
\end{figure}

Since QNC shows one-photon absorption in the visible range, the fact that it is a two-photon absorber at 772~nm is not surprising. In order to rationalize the transition from a two-photon to an apparent three-photon regime at about 15~mJ/cm$^2$, we need to consider two possible mechanisms. Three-photon absorption can take place as an instantaneous process and this can be readily observed in the wavelength ranges outside of that of two-photon absorption, where summing of the energy of three photons is needed in order to reach an excited state \cite{Samoc}. On the other hand, observation of a cubic relation between the absorption rate and the light intensity in the two-photon absorption wavelength range is very likely due to a sequential process where two-photon absorption is followed by absorption of a third photon by an excited state \cite{Zareba}. {\color{black}To determine which mechanism is operative in the present case, we carried out transient absorption experiments with a femtosecond supercontinuum pulse as the probe and a 420~nm pump that closely mimics two-photon absorption in the NIR at 772~nm, Fig.~\ref{fig4} and Supplemental. The time-resolved spectra shown in Fig.~\ref{fig4} reveal the expected ground level depletion (absorption saturation) signals below 600~nm as well as the presence of broad and strong transient absorption (excited state absorption) band from 660~nm up to 780~nm and beyond. Since the employed pump wavelength was 420~nm (producing excited states at energies close to those obtained with two photons in the NIR), it can be surmised that absorption at that wavelength promotes first the vibrationally excited, selection rules-allowed, excited state and then, in a second step, excited state absorption of a single photon can take place, leading to higher electronically excited states. With the use of NIR photons only, the process of two-photons absorption can thus be followed by a third photon absorption, at the same wavelength in the NIR, which ends up to be seen as an effective absorption of three photons.} Then, in both two-photon and three-photon intensity regimes, the excited states relaxation gives rise to heat and thermoelastic strain. Based on these time-resolved spectra measurements and on the TDBS data, we conclude that the three-photon absorption observed here is indeed a sequential process. Further insight on this sequential process would require additional time-resolved experiments that are beyond the scope of this paper and beyond the capabilities of the current spinning TDBS setup, that can only probe at a fixed probe wavelength.

Let us analytically elaborate this multiphoton scenario. The laser-excited strain amplitude derived by Zeuschner et al.\cite{Zeuschner} for two-photon absorption can be expanded in a straightforward way to include also three-photon absorption as follows,
\begin{equation}
\label{eqn:strain_nlo}
\eta_{33} = \frac{\gamma}{\rho \, C_p}\sqrt{\frac{\ln{4}}{\pi}}\left(\alpha^{(2)}\frac{F^2}{\tau}+\alpha^{(3)}\frac{F^3}{\tau^2}\right),
\end{equation}

\noindent where $\eta_{33}$ is the unidirectional longitudinal strain, $F$ the effective laser fluence, $\rho$ the film density, $C_p$ the specific heat, $\gamma$ the linear thermal dilatation coefficient, $\tau$ the FWHM laser pulse duration, and $\alpha^{(2)}$ and $\alpha^{(3)}$ the 2-photons and 3-photons absorption coefficients respectively. Note that in Eq.~\eqref{eqn:strain_nlo}, the effective fluence $F$ is linked to the laser fluence in air $F_0$ through $F \sim F_0 (1-\mathcal{R})$ with $\mathcal{R}$ being the air/film optical reflectivity at the pump wavelength. The strain given by Eq.~\eqref{eqn:strain_nlo} is directly proportionnal to the amplitude of the Brillouin signal recorded in the glass substrate. To go beyond the analytical model described in Ref.~\citen{Maris1991}, we performed Finite Element Modeling (FEM) simulations for a quantitative understanding of the excitation, propagation throughout the QNC/glass binary medium with many partial acoustic reflections at the QNC/glass boundary, as well as the detection process governing the recorded reflectivity signals, see the Supplemental for more details. Succinctly, the FEM allowed us to extract the QNC speed of sound as well as its photoacoustic coefficient, which is about tenfold stronger than that of glass. 

More importantly, the FEM revealed that the Brillouin amplitude $A_1$ matches almost perfectly the strain $\eta_{33}$ excited in QNC. This fortuitous coincidence expressed as $A_1 \equiv \eta_{33}$ makes the extraction of the nonlinear coefficients of Eq.~\eqref{eqn:strain_nlo} much simpler. Since the coefficients $\gamma$, $\rho$ and C$_p$ related to the properties of the absorbing material in Eq.~\eqref{eqn:strain_nlo} are most often known or can be measured by conventional techniques, see Supplemental, a straightforward quadratic or cubic fit of the Brillouin amplitude $A_1$ versus the pump fluence will yield the nonlinear absorption coefficients. Following this procedure, the fit displayed in Fig.~\ref{fig3} yields a two-photon absorption coefficient $\alpha^{(2)}$~=~0.74~cm/GW and a three-photon absorption coefficient $\alpha^{(3)}$~=~6.79~$\times$10$^{-2}$ cm$^3$/GW$^2$,  with a very good r-squared fitting coefficient of 0.98. As e.g. in Ref.~\citen{Liaros2017}, these nonlinear absorption coefficients can be converted into the molecular absorption cross-sections $\sigma^{(2)}$~=~6.46~GM and $\sigma^{(3)}$~=~1.52$\cdot10^{-78}$~cm$^6$.s$^2$, respectively. These values are in the ranges expected for conjugated molecules of similar size.

In light of the values of the NLO absorption coefficients determined from our TDBS measurements, we can elaborate a posteriori why the open aperture Z-scan measurements performed on the exact same sample as that for the TDBS measurements were unsuccessful, see Supplemental for details. The change in Z-scan transmittance is estimated to be well below 0.1~\%. With the typical noise level of the Z-scan technique of about 1~\%, this slight change in transmittance is clearly undetectable. On the other hand, TDBS deals with typical pump-probe signal levels that scale with the strain $\eta_{33}\equiv A_1$, typically in the range of 10$^{-3}$-10$^{-6}$, {\color{black}see the y-scale of Fig. 3(a). Undoubtedly, thanks to the benefit of the pump-probe modulation and lock-in detection scheme, the TDBS technique is orders of magnitude more sensitive than standard Z-scan and well-adapted for nanomaterials.} Another important benefit of TDBS is that the experiments can be performed in a wide range of fluences. Unlike Z-scan measurements that are limited by NLO response of the substrate holding the sample, e.g. the supercontinuum generation that interferes with the detection of nonlinear absorption, TDBS is not affected by the onset of this process.

 
We bring attention to the possibility of using slightly modified conventional time-domain Brillouin scattering measurement for the determination of multi-photon absorption coefficients of nanoscale samples. We demonstrate this capability in the case of a thin film of the organic pigment, quinacridone. We reveal the mechanism of the photoacoustic effect which includes two-photon absorption in the low fluence range, followed by a sequential absorption of a third photon in the high fluence range. The employed photoacoustic pump-probe technique is well suited for the investigation of nonlinear absorption processes in a wide variety of nanoscale solids or soft materials prepared in many different forms, amorphous, single or poly-crystalline, ordered or disordered. Our experimental results and numerical modeling lay the groundwork for future studies of ultrafast NLO processes in nanomaterials or ultra-thin films.

\section*{Supporting Information} 

Technical details (Sample, Femtosecond setup), Conventional Z-scan transmittance measurements; Finite Element Modeling of the transient reflectivity signal; X-ray diffraction (XRD) from a QNC powder sample; Analytical modeling and quantitative evaluation of the thermoelastic effect upon multiphoton absorption; Absorption spectrum, supercontinuum pump-probe measurements

\section*{Acknowledgements}

The authors acknowledge financial support from CNRS (Centre National de la Recherche Scientifique) under grant Projet International de Coop\'eration Scientifique, from Agence Nationale de la Recherche under grant ANR-16-CE30-0018, and from National Science Centre Poland (NCN) under the grant Harmonia UMO-2016/22/M/ST4/00275. This material is based upon work supported in part by the U. S. Army Research Office through the Institute for Soldier Nanotechnologies at MIT, under Cooperative Agreement Number W911NF-18-2-0048. We acknowledge Eric G\l{}owacki for sample preparation.

\bibliographystyle{apsrev4-1}

\clearpage



\section*{Supplementary material}

# Nonlinear optical absorption in nanoscale films revealed through ultrafast acoustics

Ievgeniia Chaban,[†] Radoslaw Deska,[‡] Gael Privault,[¶] Elzbieta Trzop,[¶] Maciej Lorenc,*,[¶] Steven E. Kooi,[§] Keith A. Nelson,[†] Marek Samoc,[‡] Katarzyna Matczyszyn,*,[‡] and Thomas Pezeril*,[†],[¶]

[†]Department of Chemistry, Massachusetts Institute of Technology, Cambridge, MA 02139, USA
[‡]Advanced Materials Engineering and Modelling Group, Wroclaw University of Science and Technology, PL-50370 Wroclaw, Poland
[¶]Institut de Physique de Rennes, UMR CNRS 6251, Université Rennes 1, 35042 Rennes Cedex, France
[§]Institute for Soldier Nanotechnologies, Massachusetts Institute of Technology, Cambridge, MA 02139, USA

E-mail: maciej.lorenc@univ-rennes1.fr; katarzyna.matczyszyn@pwr.edu.pl; pezeril@mit.edu

# Supplementary Information

## Technical details

**Sample** The PVD procedure has been described in Ref. 1. PVD deposition leads to a well packed film composed of many QNC clusters which form individual nanocrystals of ellipsoidal shape, as depicted in Fig. 1(b). QNC nanocrystal films can have four distinct polymorphic phases.[2] A comparison of the UV-Vis absorbance spectra of the QNC films with the spectra of each of these phases, see Refs. 3, 4 and Fig. S5, showed that the $\alpha_2$ and $\gamma$ polymorphs are predominant. The film was found to have good thickness uniformity over the entire substrate. Atomic force microscopy (AFM) and spectroscopic ellipsometry were used to extract the thickness and optical refractive indices of the film, important parameters for the Finite Element Modeling simulations.

**Femtosecond setup** Ultrafast pump-probe photoacoustic measurements were performed using a Coherent Ti-Sapphire RegA 9000 regenerative amplifier. This laser system emits a train of pulses of about 160 fs at a repetition rate of 250 kHz and at a central wavelength of 772 nm. The laser output is split into two beams. The pump beam is synchronously modulated in amplitude by an acousto-optic modulator with a 50 kHz square wave, a subharmonic frequency of the laser repetition rate. After the modulator, the pump beam travels through a delay stage which allows for control of the timing between the arrival of the pump and probe pulses. The probe beam is frequency doubled to 386 nm in a nonlinear birefringent BBO crystal ($\beta$ - $BaB_2O_4$). As illustrated in Fig. 1(c), just before being focused by a 10× microscope objective onto the sample surface, the collinear pump and probe beams get reflected by a spinning wedge mirror. This mirror, spinning at 60 Hz, is used to perform the pump-probe measurements at alternating locations on the sample surface. The pump-probe spots have Gaussian spatial beam profiles of ∼40 $\mu$m FWHM for the pump and ∼8 $\mu$m FWHM for the probe. These spots are spatially overlapped on the sample and raster across the surface in a circular trajectory of about 200 $\mu$m in diameter. After reflection from the sample, a telescope is

1

used in the probe path in order to account for the probe beam pointing variation caused by the spinning mirror and to keep the signal in the photodiode sensor area.

## Conventional Z-scan transmittance measurements

Z-scan measurements were performed employing an amplified femtosecond laser system consisting of a Coherent Astrella regenerative amplifier with a TOPAS Prime optical parametric amplifier and NIRUVis frequency mixer producing 1 kHz train of 50 fs pulses. The measurement reported here was carried out at 800 nm wavelength. A standard Z-scan setup was used (similar e.g. to that described by us in ref.[5]) to collect the transmittance of the focused laser beam through the sample in both open- and closed-aperture modes (OA and CA, respectively) simultaneously. The analysis of the data was carried out employing Sheikh-Bahae et al. theory[6] using the approach in which the light intensity in the setup is not measured directly but derived from the nonlinear phase shift determined for a sample with a known nonlinear refractive index (cf. ref.[7]), in the present case it was a 3 mm thick fused silica plate, the phase shift was 0.891 rad and the light intensity was evaluated to be 137.17 GW/cm$^2$.

Fig. S1(a) shows OA and CA traces obtained for the same sample as that investigated using the photoacoustic method, i.e. the 38 nm film of QNC deposited on a 1 mm thick soda-lime glass slide as well as the CA trace obtained for the glass slide without the deposited layer. The amplitude of the peak-to-valley distance in the CA trace of the bare glass slide corresponds to the nonlinear phase shift of 0.326 rad and for the slide with the deposited film it was 0.279 rad. For both the bare glass slide and the slide with the QNC film no obvious dip in the OA scans was observed. It should be noted that an open-aperture Z-scan can be considered as a way to perform a measurement of intensity-dependent transmittance. By converting the OA scan plotted as transmittance vs. $z$ to a transmittance vs. intensity plot (using the relation I= I$_{max}$/(1 + (z/z$_R$)$^2$)

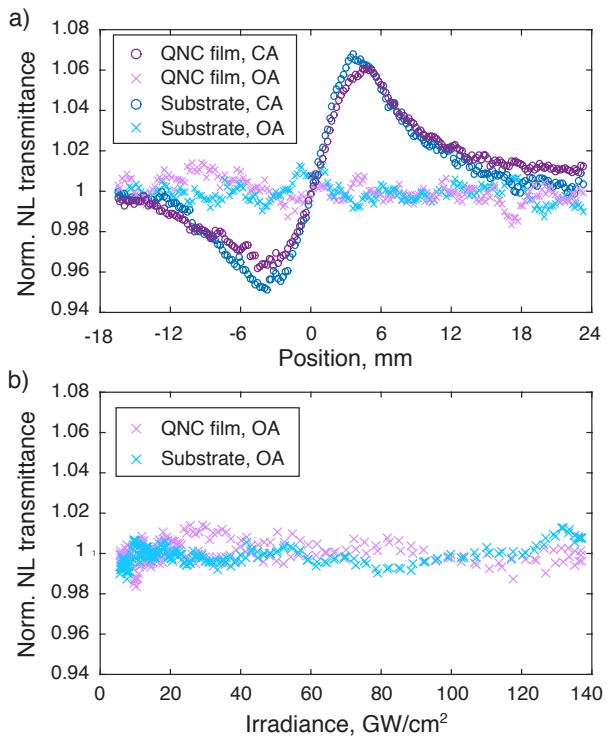

Figure S1: (a) Z-scan traces of normalized nonlinear transmittance through the QNC film coated on the soda-lime substrate and through the soda-lime substrate only. CA and OA traces reveal, respectively, the refractive nonlinearity and the nonlinear extinction. (b) Normalized nonlinear transmittance rescaled as a function of irradiance. In the whole range of irradiance, the absorption signal of the QNC film is indistinguishable from the noise even though refractive nonlinearity is observed. Within the noise range no significant trend can be extracted.

where the Rayleigh length is $z_R = \pi w_o^2/\lambda$ and $w_o$ is the beam waist at the focus, equal to 35.15 $\mu$m in the experiments described here, yielding $z_R$= 4.85 mm) one obtains the dependence shown in Fig. S1(b).

The CA data obtained for the QNC/glass sample are consistent with the notion that the nonlinear refraction quantified as the nonlinear phase shift is mostly given by the contribution of the 1 mm glass with the slight difference between the values obtained for the glass plate alone and the glass plate with the film attributable to variations in factors such as losses due to reflectance and scattering rather than the possible influence of the nonlinear refrac-

2

tive index of QNC on the overall phase shift. On the other hand, the absence of discernible nonlinear absorption dips in the OA scans is not unexpected, taking into account the relevant parameters. The nonlinear two-photon absorption coefficient for QNC derived from the photoacoustic measurement is $\alpha_2 = 0.74$ cm/GW. The transmittance $T$ of a thin film under two-photon absorption conditions is roughly given by $1/T = 1 + \alpha_2 I L$ where I is the intensity and $L$ is the sample thickness (note that this is not averaged over the Gaussian shape of the beam, but such an approximation should be sufficient for an estimate). Taking I = 100 GW/cm$^2$ and $L = 50$ nm $= 5 \times 10^{-6}$ cm, $\Delta T = \alpha_2 I L = 3.7 \times 10^{-4}$. This indicates that the change in transmittance (which is the dip in the open aperture trace) should be only 0.037 %. With the typical noise level of about 1 % this would be undectable. One should note that the intensity employed in the scans shown in Fig. S1 has been chosen in such a way that it falls in the range of the intensities used for the photoacoustic measurements, but was carefully adjusted to be below the range where either the QNC film could be damaged or the intensity was high enough to cause white light supercontinuum generation in the glass substrate.

## Finite Element Modeling of the transient reflectivity signal

To go further beyond a qualitative analytical estimate of the detected signals, we have performed Finite Element Modeling (FEM) simulations for more quantitative understanding of the excitation as well as the detection processes governing the reflectivity signals recorded in the QNC/glass bilayer. At first, we have computed time-domain FEM simulations using the k-Wave toolbox,[8] to model the one-dimensional acoustic propagation in the multilayer sample. The input parameters such as the speed of sound, the density, and the thicknesses of the different layers, used in the simulations with a chosen spatial resolution of 0.2 nm and a time step of 5 fs, are those listed in Table S1.

Since only the QNC layer absorbs the pump light through multi-photons absorption, the

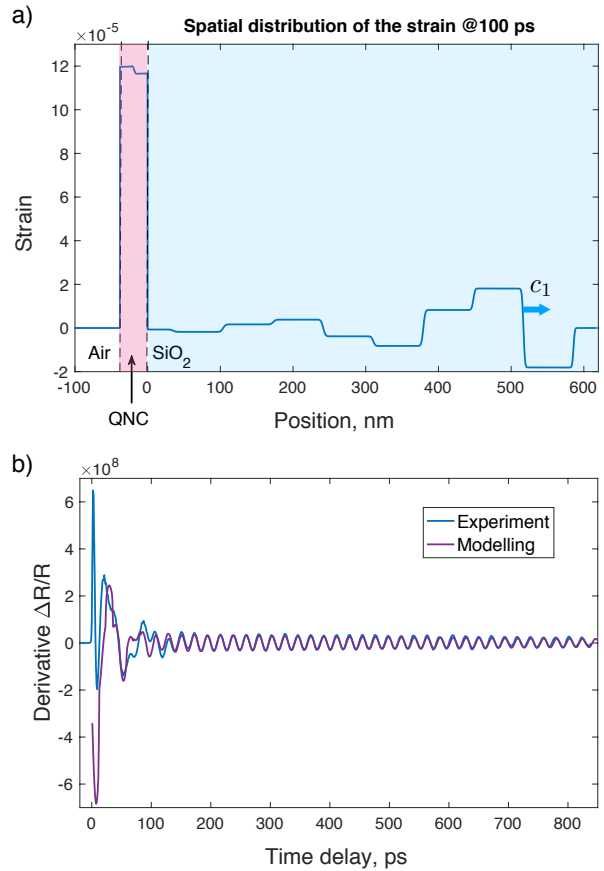

Figure S2: (a) FEM modeling of the spatial distribution of the strain at a time delay of 100 ps. (b) Comparison of the complete FEM modeling with the corresponding data taken at 27.2 mJ/cm$^2$. Apart from the onset of the laser excitation at time zero, which is out-of-scope of our FEM simulation, the simulation agrees very well with the experimental data.

laser excitation of the acoustic waves only occurs in QNC. The input acoustic source excitation was thus modeled as an instantaneous pressure jump located at these layers. Since the nonlinear absorption in QNC most probably has a characteristic depth on the order of centimeters, we can safely consider that the input pressure is uniform throughout the thickness of QNC. The k-Wave FEM solves the acoustic propagation based on the classical first-order system of equations that couples the acoustic displacements, the acoustic strain, the density and the sound speed. The calculated strain at 100 ps time delay displayed in Fig. S2(a) agrees with the expected theoretical strain profile calculated in a simplified manner in Ref. 9.

3

Table S1: Mechanical, optical and photoelastic properties of the QNC/glass bilayer sample used or extracted from the simulation.

|  | QNC film | Glass substrate |
|---|---|---|
| Density, kg/m$^3$ | 1540 | 2196 |
| Speed of sound, m/s | 3300 | 5968 |
| Refractive index | $1.549 - 0.050 \cdot i$ | 1.4876 |
| Photoelastic coefficient $\partial n / \partial S_{zz}$ | -5 | -0.5 |

As evidenced in Fig. S2(a), since the acoustic impedance of the glass substrate is much bigger than for QNC, the transmitted strain in the substrate is much weaker than in the QNC film.

As a second step, the calculated acoustic strains, at all given times and at all sample coordinates, were then used as an input to model the optical detection through time-resolved reflectivity. For the FEM modeling of the optical detection, we have used the matrix multilayer computation method to calculate the static as well as the dynamic reflection coefficients at the top interface.[10] The multilayer sample was truncated into many different optical cells of the same dimensions as for the FEM simulations of the acoustic propagation. The multilayer optical calculations solved the electromagnetic propagation at each cell boundary, based on the matrix propagation method that computes the propagation of the electromagnetic field at the boundaries. The photoelastic effect – the change in index of refraction of the optical cell due to strain acoustic perturbation – was computed in the FEM optical simulations. All the input optical parameters used in the simulation are listed in Table S1.

The full FEM simulation of the derivative of the transient reflectivity is shown in Fig. S2(b). The simulations and the data, in particular for the Brillouin scattering data in the substrate which is prevalent from 100 ps to 800 ps in Fig. S2(b), agree perfectly well. The zero time spike induced by the pump pulse is not accurately reproduced in the FEM simulation for the reason that it does not account for the light-induced ultrafast changes of the refractive index that occur during the onset of acoustic excitation. Importantly, since the photoelastic coefficient of glass at the probe wavelength is well known,[11] the FEM simulation can be used to extract the laser-excited input strain in QNC at zero time. As shown in Fig. S2(a), at the highest fluence of 27.2 mJ/cm$^2$, the input strain in the QNC film is of $12 \times 10^{-5}$, which is relatively weak and indicates that the nonlinear Brillouin signal is not related to nonlinear acoustics that could appear at substantial strain amplitudes.[12] The nonlinear fluence dependence is undoubtedly linked to nonlinear optical absorption. The FEM simulations were used as well to extract the photoelastic coefficient of QNC at the probe wavelength, which appears to be about tenfold stronger than the one for glass, see Table S1. The accuracy of the quantitative extraction of the QNC photoelastic coefficient is related to the reflectivity fit in the 10-100 ps time delay range for which the QNC optical response dominates, see Fig. S2(b). The high value of the extracted QNC photoelastic coefficient could have important practical applications in the design of efficient and ultrafast photoelastic modulators.

## X-ray diffraction (XRD) from a QNC powder sample

To study the evolution of the unit-cell parameters with temperature and to quantitatively extract the concentration of the polymorphic phases in the studied QNC sample, powder XRD data were collected on an Agilent Technologies SuperNova Single-Crystal X-ray diffractometer with a micro-source, using Cu-K$\alpha$ radiation ($\lambda = 1.54$ Å), and fitted with an EosS2 detector. The QNC sample was measured in a borosilicate glass capillary of 0.3 mm outer diameter. A nitrogen flow 800Plus series cryostat from Oxford Cryosystems was used for the measurements with temperature varied in cooling mode from 320 K to 100 K with a 10 K

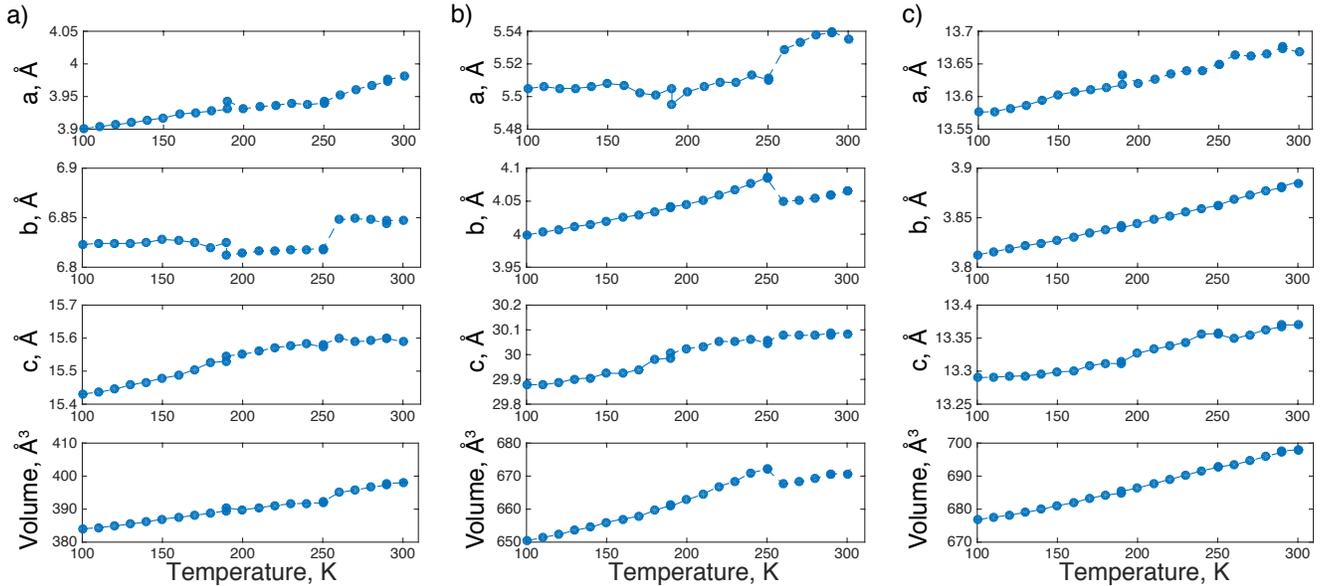

Figure S3: Unit-cell parameters change with temperature respectively for $\alpha$- (a), $\beta$- (b) and $\gamma$- (c) phase in a multi-phase QNC sample.

step. At each new temperature, the sample was given additional 5 min to allow the temperature to equilibrate within the capillary volume. The QNC sample was 360° spun around the $\phi$-axis for 4 min during data collection. The X-ray detector was placed at 55 mm distance from the sample. As shown in Fig. S4, the measured diffraction patterns from the QNC sample shift with temperature.

In order to refine the X-ray diffraction peaks obtained at different temperatures, we have used the CrysAlisPRO[13] software package from Rigaku Oxford Diffraction. Rietveld refinement was carried out in TOPAS.[14] The refinement was performed in the wavevector range from 0.40 to 3.18 $Å^{-1}$, corresponding in this case to diffraction angles from 5.5° to 46°. The background was described with a Chebyshev polynomial. Structural information including atomic coordinates was imported into TOPAS for all three phases of QNC ($\alpha$, $\beta$ and $\gamma$). The models used here were previously reported by Paulus et al.[2] (Cambridge Structural Database reference numbers: CCDC620257-CCDC620259). Stability of the refinement was assured by restraining the molecular geometry of QNC within a rigid body defined on a half molecule, i.e. an asymmetric unit, and fixed to an inversion center with a dummy atom placed respectively at (0,1/2,0) for $\alpha$-phase, and at (1/2,0,0) for $\beta$- and $\gamma$-phases.

During refinement, the only free parameters were the unit-cell parameters, the peak profile parameters, the scale factors and the rigid body position for each of the three phases. In order to retain the simplest convolution functions and to prevent the divergence of the refinement, no Lorentzian contribution and anisotropy was considered. Regardless of the limited wavevector region and rather poor Q-resolution, the Rietveld refinement for each temperature was quite reasonable (with an average R-factor not exceeding $\sim$2.7 %). The Rietveld fit confirmed the presence of three polymorphic phases in the

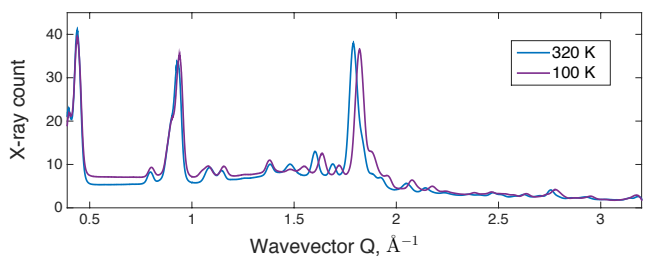

Figure S4: Powder x-ray diffraction patterns of a multi-phase QNC sample taken at 320 K and 100 K respectively. The shift in wavevector Q ($Å-1$) indicates the system response to temperature.

5

studied QNC sample. An averaged composition was estimated respectively to 5.3 % of $\alpha$-phase, 23.2 % of $\beta$-phase and 71.5 % of $\gamma$-phase. The predominance of the $\gamma$ phase revealed by these X-ray diffraction analyses is in agreement with the comparison of the QNC film absorbance spectrum with the absorbance spectrum of each of the phases.[3,4] Further analysis also revealed the unit cell parameters change with temperature for each of the three QNC phases, see Fig. S3.

## Analytical modeling and quantitative evaluation of the thermoelastic effect upon multiphoton absorption

The most common mechanism of generation of longitudinal strain waves by light that occurs in absorptive materials is the well-known thermoelastic effect.[15] In this case, the absorption of the light energy produces a transient temperature rise that scales linearly with the amount of absorbed energy that is converted to heat. This sudden temperature jump causes lattice distortion through thermal dilatation that, in the case of a thin film, primarily drives the acoustic excitation of unidirectional strain waves $\eta_{33}$ that propagate along the surface normal $x_3$. Assuming that the generated heat is proportional to the amount of energy absorbed through two-photon and three-photon processes, one can postulate that the generated strain amplitude is given by Eq. (3) in the main text, which is similar to that given in Ref. 16 but it involves the linear and not volumetric thermal dilatation coefficient, as well as inclusion of three-photon absorption alongside that of two photons. An important detail included in Eq. (3) is the fact that the laser fluence $F$ is the effective laser fluence that gets absorbed in the film, which is different in comparison to the fluence in air $F_0$ by reflection at the air/film interface and at the film/substrate interface. The optical reflectivity at the air/film interface can be calculated from the QNC refractive index listed in Table 1, to be $\mathcal{R} = 0.21$. On the other hand, the optical reflectivity at the QNC/glass interface can be calculated to be only 2%, so any Fabry-Perot cavity effects can be neglected and then the effective fluence is simply $F \sim F_0 (1 - \mathcal{R})$.

Overall, from the measured temperature evolution of the different cell parameters of the phases in Fig. S3, and the determination of the different phases content in percentage, we can calculate the effective thermal dilatation coefficient of the QNC film which is needed in order to estimate the nonlinear optical absorption coefficients. This effective thermal dilatation coefficient is calculated from the estimate of the thermal dilatation coefficients of the different phases weighed by their percentage. For our specific QNC sample, we can calculate the effective thermal volume expansion coefficient to be $1.71 \times 10^{-4}$ K$^{-1}$. However, since the strain excited along the normal of the sample surface should be treated as unidirectional, the thermal dilatation coefficient needed for the strain calculation is the linear thermal dilatation coefficient which, on the average, can be taken as simply 1/3 of the volumetric thermal dilatation coefficient, giving the value of $\gamma = 57.1 \times 10^{-6}$ K$^{-1}$. The specific heat $C_p$ of QNC if not available in the literature, but it should not be very different from the heat capacity of a very similar molecule of pentacene, thus we assume the heat capacity of QNC at room temperature to be $C_p \sim 1100$ J.kg$^{-1}$.K$^{-1}$, see the table in Ref. 17.

Finally, upon determination of all the unknown parameters of Eq. (3) in the main text, such as $\gamma$, $\rho$ and C$_p$, a numerical quadratic or cubic fit of Fig. 3 in the main text – that displays in the A$_1$ amplitude ($\equiv$ strain $\eta_{33}$) versus the laser fluence in air – lead to the determination of the nonlinear $\alpha^{(2)}$ and $\alpha^{(3)}$ coefficients.

## Absorption spectrum, supercontinuum pump-probe measurements

As shown in Fig. S5, UV-Vis absorption spectrum of QNC film is dominated by a transition peaking at about 550 nm (having the character of intramolecular charge transfer). One-

6

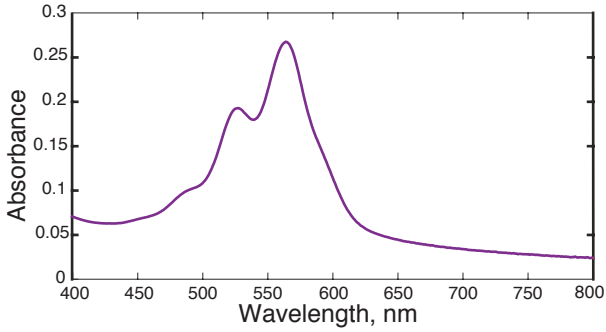

Figure S5: Absorbance spectrum of a 37 nm thick QNC film on a glass substrate. The non-zero offset of absorption, clearly visible at long wavelengths, is attributed to reflectivity and scattering losses.

photon absorption at 772 nm is negligible and the excitation at this wavelength is only possible through multiphoton processes. It should be noted that the molecule of QNC has a centre of symmetry and therefore one should expect that different excited states are reachable according to selection rules for one- and two-photon transitions, as depicted in Fig. S6 (b). Namely, one-photon transitions are allowed for a change in parity (g $\rightarrow$ u), whereas two-photon transitions should preserve the parity (g $\rightarrow$ g). The selection rule for an instantaneous three-photon transition should be again (g $\rightarrow$ u). Therefore, while one-photon excitation at 550 nm leads to formation of the $S_1$ state, it is expected that excitation at 772 nm proceeds through a two-photon absorption process reaching the second excited state $S_2$. With high fluence excitation in NIR, one cannot exclude the possibility of a coexistence of an instantaneous three-photon absorption in a range of wavelengths, but cubic fluence dependence can also result from a sequential scenario whereby two photons generate an excited state (which in the case of a molecular solid can be deemed to be a Frenkel exciton) and absorption of a third photon by the exciton. To substantiate a sequential three-photon process we carried out femtosecond supercontinuum transient absorption experiments.

The pump wavelength was set to 420 nm (with FWHM of 10 nm) for convenience by use of an Optical Parametric Amplifier (TOPAS, Light Conversion, pumped by a regenerative

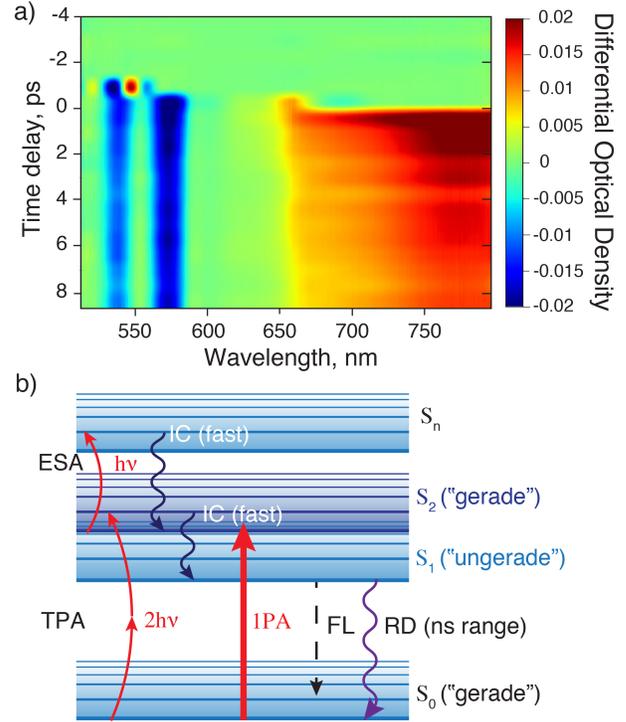

Figure S6: (Color online) (a) Transient linear absorption of QNC nanocrystals at ultrafast timescale for a 420 nm pump: red - excited state absorption, blue - ground state depopulation. Spectra around time zero were not corrected for chirp; (b) Tentative excitation and relaxation diagram: TPA - two-photon absorption (772 nm), 1PA - one-photon absorption (420 nm), IC - internal conversion, FL - fluorescence, RD - non-radiative decay.

amplifier operating at 1 kHz, Legend USP Coherent). The $\sim$100 femtosecond pump was focused to a 200 $\mu$m diameter spot on the sample. The pulse energy was kept in the range 0.5-1.5 $\mu$J. The pump beam was modulated at 500 Hz with an optical chopper to allow on/off sequence for the probe and improved signal-to-noise. The probe was a femtosecond supercontinuum pulse (450 - 850 nm) generated with a 1310 nm pulse from a second OPA and focused on a sapphire plate of 3 mm thickness. The supercontinuum probe beam was focused to 100 $\mu$m spot on the sample with a silver parabolic mirror. The spectra from the sample were dispersed with a monochromator (Acton SP2500) and recorded with a Basler acA2440-20gm GigE camera at 1 kHz frequency.

The time-resolved spectra shown in Fig. S6 (a)

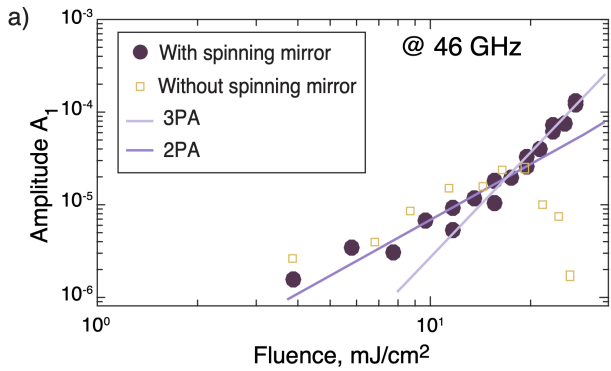

Figure S7: (Color online) Fluence dependence of the 46 GHz Brillouin oscillations, replica of Fig 3(a) in the main text.

reveal expected ground level depletion (absorption saturation) signals that occur at wavelengths corresponding to those of 1PA peaks as well as the presence of broad and strong transient absorption (excited state absorption) from 660 nm up to 780 nm and beyond. Since the employed pump wavelength was 420 nm, it can be surmised that absorption at that wavelength promotes first the vibrationally excited symmetry allowed excited state, $S_1$ (*ungerade*), which is energetically quite close to the excited state reachable by two-photon absorption in the NIR, $S_2$ (*gerade*). Excited state absorption can take place from both $S_1$ and $S_2$, leading to higher electronically excited states $S_n$.

In the case of the photoacoustic measurements, given a relatively long lifetime of the $S_1$ state (nanoseconds[18]), we argue that the main contribution to the strain build-up comes from the ultrafast internal conversion of the $S_n$ states as well as vibrational relaxation of the $S_2$ and $S_1$ states. The former contributes predominantly to strain at higher laser fluence, showing cubic dependence, the latter contribute to strain at lower fluence, where assuming two-photon absorption only provides satisfactory fits, see Fig. S7 which is a replica of Fig. 3(a) in the main text.



\end{document}